\documentclass[aps,prb,twocolumn,superscriptaddress,amsmath,nofootinbib,10pt,floatfix]{revtex4-2}
\usepackage{graphicx}
\usepackage[normalem]{ulem}
\usepackage{xcolor}
\usepackage{textgreek}
\usepackage[colorlinks, linkcolor=blue, citecolor=blue]{hyperref}
\usepackage{gensymb}

\usepackage{amssymb}
\newcommand{\be}{\begin{equation}}
\newcommand{\ee}{\end{equation}}
\newcommand{\beq}{\begin{eqnarray}}
\newcommand{\eeq}{\end{eqnarray}}
\newcommand{\ba}{\[\begin{aligned}}
\newcommand{\ea}{\end{aligned}\]}

\newcommand{\la}{\langle}
\newcommand{\ra}{\rangle}

\renewcommand{\vec}[1]{{\bf #1}}
\renewcommand{\epsilon}{\varepsilon}

\def\nn{\nonumber}

\newcommand{\eqn}[1]{\begin{equation} #1 \end{equation}}
\newcommand{\eqa}[1]{\begin{align} #1 \end{align}}
\newcommand{\avg}[1]{\left\langle #1 \right\rangle}

\renewcommand{\vec}[1]{\boldsymbol{#1}}

\def \bS{{\bf S}}

\def \bq{{\bf q}}

\def \W{{\Omega}}
\def \e{{\epsilon}}
\def \ve{{\varepsilon}}

\def \w{{\omega}}

\def \k{{\vec{k}}}
\def \q{{\vec{q}}}
\def \K{{\vec{K}}}

\def \e{{\epsilon}}

\def \S{{\mathcal{S}}}

\def \ra{{\rangle}}
\def \la{{\langle}}
\def \tn{\textnormal}

\def \ba{\begin{align*}}
\def \ea{\end{align*}}

\DeclareMathOperator{\arcsinh}{arcsinh}
\newcommand{\Z}{\mathcal{Z}}
\newcommand{\J}{\mathcal{J}}
\newcommand{\mD}{\mathcal{D}}

\newcounter{indice}

\newcommand*{\DC}[1]{\textcolor{orange}{ DC: {#1}}}

\newcommand*{\amcr}[1]{\textcolor{red}{ AMcR: {#1}}}
\begin{document}

\title{Intermediate-scale theory for electrons coupled to frustrated local-moments}
\author{Adam J. McRoberts}\thanks{These authors contributed equally}
\affiliation{Max-Planck-Institut f\"ur Physik komplexer Systeme, N\"othnitzer Stra\ss e 38, 01187 Dresden, Germany}
\author{J.F. Mendez-Valderrama}\thanks{These authors contributed equally}
\affiliation{Department of Physics, Cornell University, Ithaca, New York 14853, USA.}
\author{Roderich Moessner}
\affiliation{Max-Planck-Institut f\"ur Physik komplexer Systeme, N\"othnitzer Stra\ss e 38, 01187 Dresden, Germany}
\author{Debanjan Chowdhury}
\affiliation{Department of Physics, Cornell University, Ithaca, New York 14853, USA.}
\begin{abstract}
A classic route for destroying long-lived electronic quasiparticles in a weakly interacting Fermi liquid is to couple them to other low-energy degrees of freedom that effectively act as a bath. We consider here the problem of electrons scattering off the spin fluctuations of a geometrically frustrated antiferromagnet, whose non-linear Landau-Lifshitz dynamics, which remains non-trivial at all temperatures, we model in detail. At intermediate temperatures and in the absence of any magnetic ordering, the fluctuating local-moments lead to a  non-trivial angular anisotropy of the scattering-rate along the Fermi surface, which disappears with increasing temperature, elucidating the role of  ``hot-spots". Over a remarkably broad window of intermediate and high temperatures, the electronic properties can be described by employing a {\it local} approximation for the dynamical spin-response. This we contrast with the more familiar setup of electrons scattering off classical phonons, whose high-temperature limit differs fundamentally on account of their unbounded Hilbert space.  We place our results in the context of layered magnetic delafossite compounds. 
\end{abstract}

\maketitle

{\it Introduction.-} 
Electronic solids provide a fascinating experimental platform for studying the properties of an electronic fluid coupled to a ``bath". A paradigmatic example is the coupled electron-phonon problem, where the phonons effectively act as a bath with which the electrons exchange energy and momentum \cite{ziman}. The transport and single-particle properties for numerous metals at intermediate temperatures can be understood in terms of this ``semi-quantum" system over $\omega_{\tn{D}}\lesssim T\ll \ve_{\tn{F}}$, where $\ve_{\tn{F}}$ is the electronic Fermi energy and $\omega_{\tn{D}}$ is a characteristic Debye frequency. A number of recent developments in moir\'e materials \cite{moirerev,Mak2022,FaiMIT,pasupathyMIT,ajesh,ruhman} and magnetic delafossites \cite{Mackenzie_2017, takatsu2010anisotropy, ong2012three, glamazda2014collective, takatsu2014magnetic, noh2014direct, daou2017unconventional, sun2019magnetic, komleva2020unconventional, Sunko, sunko2020controlled} inspire us to examine the following question: What is the nature of a weakly-correlated electronic fluid coupled to interacting local-moments for $J\lesssim T\ll \ve_{\tn{F}}$, where $J$ is a characteristic antiferromagnetic exchange energy scale? While this system exhibits a familiar resemblance to the electron-phonon problem, there are a number of important conceptual differences. 

First and foremost, spins have a bounded Hilbert space. While phonon modes tend to obey classical equipartition at high $T\agt \omega_\tn{D}$, spin excitations at $T\agt J$ do not. Secondly, the dynamical correlations in an interacting (``cooperative") paramagnet at $T\gtrsim J$ are {\it a priori} non-trivial, arising from a nonlinear dynamics, even though the static correlations vanish asymptotically at high temperatures. Finally, residual short-range order, reflecting any magnetic order at $T<T_{\tn{N}}~(\lesssim J)$,  can leave an imprint on the electronic properties even once the order melts at $T\gtrsim J$.

In this letter, we focus specifically on the case of local-moments with geometrically frustrated interactions on the triangular lattice \cite{chalker2017spin}. A frustrated magnet is a useful starting point as a bath, since the tendency towards any long-range magnetic ordering is naturally suppressed, 
providing a broad paramagnetic regime, which always has a non-trivial (non-linear) dynamics, with a momentum dependence reflecting, e.g., spin diffusion. Even at the highest temperatures, the question naturally arises whether this can impart non- (or marginal-)Fermi liquid-like electronic correlations \cite{Varma,hartnoll2022}. Notably, by carrying out a detailed analysis of the non-linear spin-dynamics and its effect on the electronic properties, the numerically evaluated electron self-energy can be captured quantitatively over a broad energy window $J\alt T\leq \infty$ by employing a local approximation for the spins.

{\it Model.-} We consider a simple two-dimensional model of itinerant spinful electrons, $c_{\k\sigma}$, interacting with spin$-\frac{1}{2}$ local moments, $\bS_i$, on the sites of a triangular lattice:
\begin{subequations}
\beq
H &=& H_c + H_S + H_K,\\
H_c &=& \sum_{\k,\sigma} (\ve_\k - \mu) c^\dagger_{\k\sigma} c^{\phantom\dagger}_{\k\sigma},~~
H_S = J\sum_{\la i,j\ra} \bS_i\cdot\bS_j,\\
H_K &=& g\sum_{i,\alpha,\beta} c_{i\alpha}^\dagger (\bS_i\cdot\vec{\sigma}_{\alpha\beta})c_{i\beta},
\eeq
\end{subequations}
where $\ve_\k,~\mu$ represent the dispersion and chemical potential associated with the $c-$electrons, $J(>0)$ denotes the antiferromagnetic exchange interaction between local moments, and $g(>0)$ is the Kondo-coupling between the local moment and electron spin-density, respectively. For the electronic dispersion, we include first ($t$) and second ($t'$) neighbor hoppings on the triangular lattice. We note at the outset that in the temperature window of interest and for a ``weak" Kondo-coupling, the intrinsic quantum mechanical Kondo-physics associated with the quenching of the local moment will be irrelevant. 

The dynamics of frustrated magnets in their cooperative paramagnetic phase tends to be well-described by the classical Landau-Lifshitz equations of motion governed by $H_S$, even for small spin lengths $S$ far from the classical limit $S\rightarrow\infty$ \cite{Keren_1994,Moessner_1998L,Moessner_1998B,conlon,Anjana_2017,Mourigal_2019,Shu_2019,Knolle_2022T}:
\be
\dot{\bS}_i = \frac{\partial H_S}{\partial \bS_i} \times \bS_i.
\label{eqn:ll}
\ee
We hence approximate the spins as $O(3)$ vectors precessing around their local exchange fields, which we analyse by performing molecular dynamics (MD) simulations. We average over initial configurations obtained from classical Monte-Carlo (MC) simulations of $H_S$ and numerically integrate the equations of motion \cite{si}.

In this letter, we  analyze the extent to which the dynamical correlations associated with the fluctuating local moments leave an imprint on the electronic liquid at intermediate energy scales. Specifically, we calculate the $O(g^2)$ perturbative correction to the electron self-energy, which is controlled by the two-point correlator of the spins. Since the primary goal is to understand the electronic properties, we ignore the electron back-action on the local moments, which affects the electronic properties at higher order in $g$. Previous work \cite{maslov} has analyzed this problem in a different regime, focusing on the elastic transport lifetimes.

{\it Spin dynamics.-} 
In momentum space, the two-point correlator of interest is  the dynamical structure factor,
\beq
\S(\q, \w) = \int_{-\infty}^{\infty} dt ~e^{i\w t} \avg{\bS(\q, t)\cdot \bS(-\q, 0)}.
\eeq

\begin{figure}
    \centering
    \includegraphics[width=0.475\textwidth]{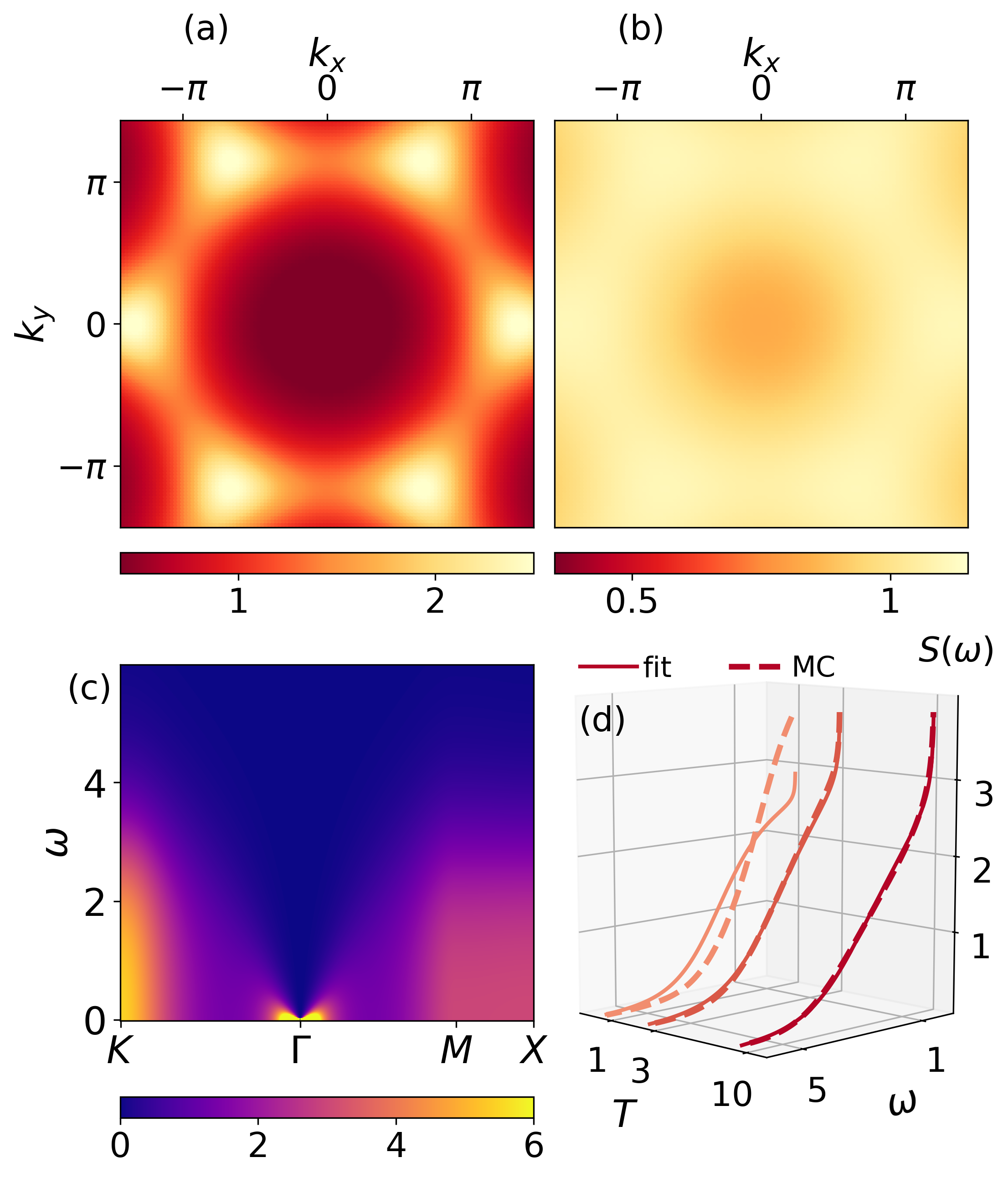}
    \caption{The static structure factor $\S(\q)$ over the Brillouin zone for (a) $T = J$, and (b) $T = 10 J$, respectively. (c) $\S(\q, \w)$  over a momentum-cut $K \rightarrow \Gamma \rightarrow M \rightarrow X$, at $T = 3J$. (d) The momentum-integrated structure factor for selected temperatures, together with the fit in Eqn.~\ref{eq:dsf_fit}. 
    }
    \label{fig:dsf}
\end{figure}

We use heatbath Monte-Carlo \cite{loison2004canonical} to draw an ensemble of 1000  independent initial states $\bS_i(0)$ from the canonical ensemble of $H_S$ at a given temperature $T$; from here on we measure $T$ in units of $J$.  
The time evolution of each state is obtained by numerical integration of Eq.~(\ref{eqn:ll}), using the standard fourth-order Runge-Kutta procedure, to a final time of $t_f = 4096J^{-1}$. We construct each Fourier transformed state $\bS(\q, \w)$, whose ensemble average  yields the dynamical structure factor. The numerically evaluated static structure factor, $\S(\q)=\int d\omega~ \S(\q,\omega)$, is shown for $T=J$ and $T=10J$ in Figs.~\ref{fig:dsf}(a) and (b), respectively. The broadened ``Bragg-like" peaks at $T=J$ near the high-symmetry points represent a remnant of the thermally disordered $120^{\degree}$ antiferromagnetic state on the triangular lattice. 

We are interested in the regime where the spin bandwidth --- the range of frequencies over which $\S(\q, \w)$ has support --- is much smaller than the electronic bandwidth and Fermi energy. 
As we discuss later, the energy scales that are relevant for electrons with momentum $\k$ and frequency $\omega$ scattering off the spins enter the structure factor as $\S(\q, \w - \e_{\k + \q})$; the latter has support only over a very narrow (and $\k$-dependent) region of momentum $\q$. Even for relatively large system sizes ($N = L^2, L = 120$), the momentum resolution available from the numerical simulations directly is insufficient to determine the electronic lifetime. We  therefore construct an analytical fit to $\S(\q, \w)$ from our numerics.

We obtain the static (equal-time) structure factor from a ``soft spin" approximation \cite{si}. We can then describe the numerically computed dynamical structure factor, for $T \gtrsim J$, using the following phenomenological {\it ansatz}:
\beq
\S(\q, \w) = \frac{\S(\q)\mathcal{N}(\alpha_\q, \eta_\q)}{\sinh^2(\alpha_\q \w) + \eta_\q},
\label{eq:dsf_fit}
\eeq
where $\alpha_\q$ and $\eta_\q$ are momentum-dependent fitting parameters, and $\mathcal{N}$ is a normalisation factor enforcing $\S(\q) = \int d\w ~\S(\q, \w)$. As both $\alpha_\q,~\eta_\q$ respect the  space group symmetries of the triangular lattice, they can be expressed in terms of the following objects: 
\beq
\gamma_n(\q) = \frac{1}{|E_n|} \sum_{\vec{\delta} \in E_n} e^{i\q\cdot \vec{\delta}},
\label{eq:gn}
\eeq
where $E_n$ is the set of $n^{\tn{th}}$ nearest-neighbour vectors, and $|E_n|$ its cardinality. At high temperature, with a short correlation length, typically the first few $\gamma_n$ are sufficient to describe these fit functions. We show $\S(\q,\omega)$ along a certain high-symmetry cut in the Brillouin zone at $T=3J$ in Fig.~\ref{fig:dsf}(c). In Fig.~\ref{fig:dsf}(d) we compare our analytical fit functions to the momentum-integrated structure factors for three different temperatures. As expected, the quality of our fits improve with increasing temperature; the largest disagreement can be seen at $T=J$.

\begin{figure*}[htbp!]
\centering
\hspace*{-0.9cm} 
\includegraphics[width=1.07\textwidth]{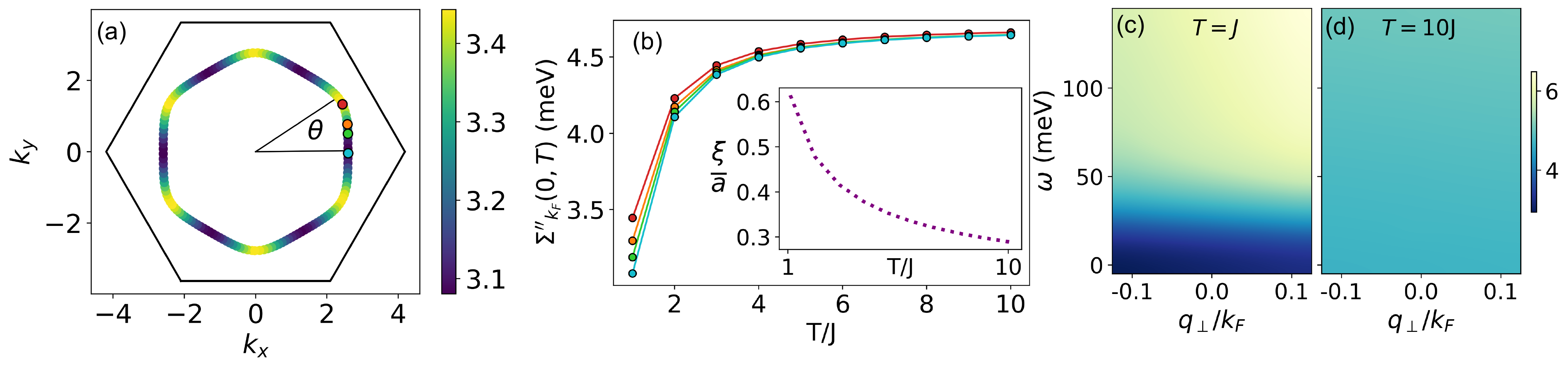}
\caption{The numerically evaluated electron self-energy, $\Sigma ''_{k_F}(0,T)$ (a) along the Fermi surface at $T=J$, and (b) for different $\theta$ along the Fermi surface as a function of temperature. Inset: The spin correlation length, $\xi(T)$, extracted from $\S(\q)$. (c)-(d) With increasing temperature, $\Sigma''_{k_F}(q_\perp,\omega)$ becomes featureless as a function of both $\omega$ and $q_\perp$.} 
\label{fig:N1}
\end{figure*}

{\it Electron self-energy.-} The imaginary part of the electron self-energy at real frequencies is given by,
\begin{subequations}
\beq
   && \Sigma''(\k,\omega)= \label{eqn:spectral_rep_sigma}\\ 
  &&  \frac{g^{2}}{N}\sum_{\q}\int\frac{d\Omega}{\pi}\ \chi_{\tn{spin}}''\left(\q,\omega-\epsilon_{\k+\q}\right)A_c\left(\k+\q,\Omega\right)f(\omega,\Omega),\nn\\
  && f(\omega,\Omega) = \left[n_{b}\left(\omega-\Omega\right)+n_{f}\left(-\Omega\right)\right] \label{f},
\eeq
\end{subequations}
where $A_c(\k, \omega)$ is the electron spectral function, $\chi_{\tn{spin}}''(\k, \omega)$ denotes the imaginary part of the spin susceptibility, and $n_b(...),~n_f(...)$ denote the Bose-Einstein and Fermi-Dirac distributions at temperature $T=\beta^{-1}$, respectively. In the temperature window of interest, the susceptibility is related to the structure factor discussed earlier,
\beq
     \chi_{\tn{spin}}''\left(\q,\omega\right)=\beta\omega~\S\left(\q,\omega\right),
    \label{eqn:FDtheo}
\eeq
where we have used the high-temperature (``classical") version of the fluctuation-dissipation theorem for the sake of internal consistency \cite{si}. {In what follows, a key new ingredient in our computation is associated with a detailed microscopic modeling of the intermediate-scale non-linear dynamics of the spin system, which can modify the electron self-energy in interesting ways.}

We begin by noting that, despite the small spin-bandwidth, high-energy electrons ($\omega\gg J$) can scatter off the spins as long as $|\omega-\epsilon_{\k}|\lesssim J$. We first evaluate the self-energy numerically \cite{si}. In what follows, we fix $t=568$ meV, $t'=-108$ meV, $J=10.34$ meV, $g=2.5$ meV, and $\mu=247.5$ meV. These values are partly inspired by the magnetic delafossite PdCrO$_2$ \cite{Sunko}, which consists of alternately stacked layers of triangular lattice antiferromagnetic Mott insulators and weakly correlated metals near half-filling. The electronic transport properties in this compound are clearly affected by the presence of thermally fluctuating local moments \cite{Hicks15}, which necessitates a theoretical investigation of the single-particle lifetime in this unusual regime.

{\it Numerical results.-} The results for $\Sigma''_{k_F}\left(\omega=0,T\right)$ around the Fermi surface with $|\k|=k_F$ for a number of angles, $\theta$, and over a range of finite temperatures is shown in Fig.~\ref{fig:N1}(a)-(b). {There are six bright spots at $T\gtrsim J$, that we associate with hot-regions arising from scattering off short-ranged magnetic fluctuations with a finite correlation length $\xi(T)$, peaked near the $\K,~\K'$ points in the BZ. These regions are comprised of twelve ``hot-spots" identified by the condition $\ve(\k\pm \K)=\ve(\k)$  (similarly for $\K'$), that become thermally smeared into the six spots \cite{si}.}  

The behavior is reminiscent of fluctuation effects involving electrons scattering off short-ranged density-wave fluctuations in the context of a Peierls transition in one dimension \cite{LRA}. Furthermore, the angular variation of the evaluated self-energy is closely tied to the filling, and the magnitude of the $2k_{\tn{F}}$ vector relative to the ordering wavevectors \cite{si}. With increasing temperature, the angular anisotropy near the hot-spots disappears as the correlation length decreases. Ultimately, with increasing $T$, and in contrast to electrons scattering off high-temperature phonons, there appears a uniform temperature-independent saturation value for the self-energy along the Fermi surface associated with the asymptotic limit $J\ll T (\ll \ve_F)$; see Fig.~\ref{fig:N1}(b). However, this saturation sets in gently, with $\Sigma''_{k_F}\left(\omega=0,T\right)$ varying only within $\approx 20\% $ of its saturation value between $T=J$ and $T=10J$. We extract the spin correlation length, $\xi(T)$, from the real space static structure factor as a function of increasing temperature, and find that it is already smaller than the lattice spacing at $T=J$.

Next, we evaluate the frequency and (transverse) momentum dependence of the self-energy, $\Sigma''_{k_F}(q_\perp,\omega,T)$, away from the Fermi surface. In Fig.~\ref{fig:N1}(c)-(d), we show a color-map for the self-energy for a fixed $\theta$ at two different temperatures; the scales are chosen such that $\omega$ is comparable to $v_F q_{\perp}$. At $T\sim J$, the self-energy exhibits a weak $q_\perp$ dependence; the interesting feature is tied to the $\omega-$dependence for the full range of $q_\perp$ considered in Fig.~\ref{fig:N1}(c). At a fixed $q_{\perp}$, $\Sigma''_{k_F}(\omega)\sim \omega^2$ at low frequencies, and crosses over into a distinct regime with weak $\omega-$dependence for $\omega \gtrsim 2J$. With increasing temperature,  Fig.~\ref{fig:N1}(d), the self-energy becomes largely momentum independent, signaling a predominantly ``local" character of the spin-fluctuation spectrum. In such a regime, the self-energy displays a nearly featureless behavior as a function of $\omega,~q_\perp$, that varies weakly with temperature. The electronic properties for $3J\lesssim T\ll\ve_F$ can be captured by the high-temperature and local limit of $\chi''_{\tn{spin}}$, as we discuss below. 

\begin{figure}[ht!]
\centering
\includegraphics[width=0.5\textwidth]{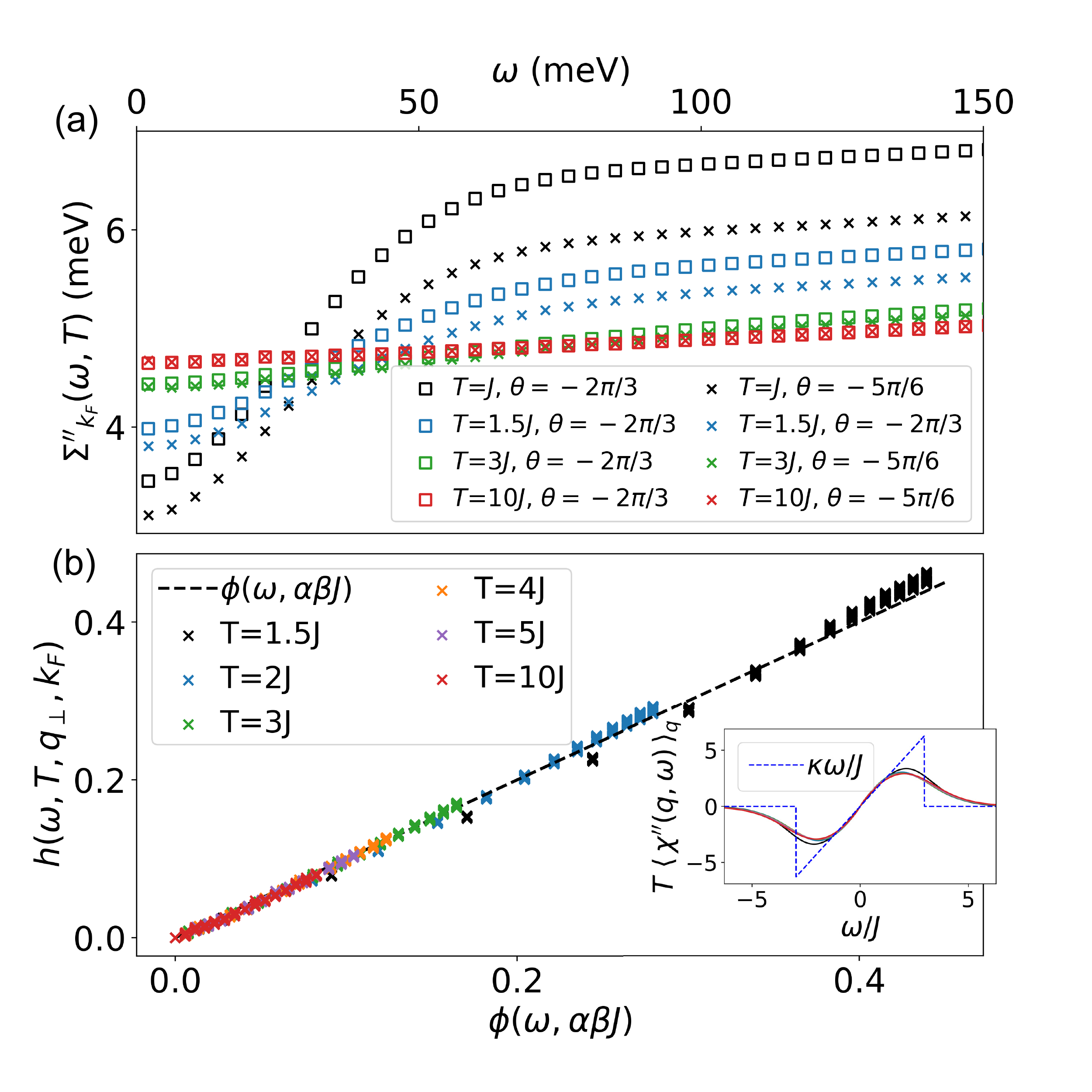}
\caption{a) Frequency dependence of the self-energy at different angles along the Fermi surface with increasing temperature. b) The function $h(...)$ constructed out of the electron self-energy (Eq.~\ref{eqn:freqdep}) exhibits a scaling collapse on a universal momentum independent curve  $\phi(\omega,\alpha\beta J)$ over a wide range of $T$, angle around the Fermi surface and $q_\perp$.} 
\label{fig:N2}
\end{figure}

In Fig.~\ref{fig:N2}(a), we analyze the $\omega$ and $T$ dependence of $\Sigma''_{k_F}$ for a range of $\theta$ along the Fermi surface. At low $T$, there is a crossover from a $\omega^2$ to a weak $\omega-$dependence at larger $\omega$; the dispersive structure disappears with increasing $T$. Despite the angular anisotropy in $\Sigma''_{k_F}(\omega=0,T)$ for small $T$, the  relative renormalization of the self-energy as a function of $\omega$ is largely insensitive to $\theta$ at a given $T$. We have also observed that even for finite $q_\perp$, the $\omega,~T$ dependencies are broadly similar to the $q_\perp=0$ results. In spite of the complex structure of the dynamical spin-response (Fig.~\ref{fig:N1}), we will be able to capture most of the {\it quantitative} features of the electron self-energy starting from a high-temperature analytical perspective. 

{\it Analytical approach for electron self-energy.-} Recall that, in the Lehmann representation,
\beq
    &&\chi''_{\tn{spin}}\left(\q,\omega\right)= \label{eqn:lehmann}\\
    &&\pi\left(e^{\beta\omega}-1\right)\frac{1}{Z}\sum_{n,m,\alpha}e^{-\beta E_{m}}\left|\langle n|S_{\q}^{\alpha}|m\rangle\right|^{2}\delta\left(\omega+E_{n}-E_{m}\right),\nn
\eeq
where the $\{n,~E_n\}$ label the many-body eigenstates and eigenenergies, respectively, and $Z$ is the partition function. The spin-operators $S_{\q}^\alpha$ denote the $\alpha-$component with momentum, $\q$.  
{Remarkably, we have observed that the numerical computation of the electron self-energy based on the full $\S(\q,\omega)$ agrees almost perfectly with a completely ``local" approximation for the susceptibility (to be made precise below). Specifically, this implies that the frequency and angular dependence along the Fermi surface of the electron self-energy is controlled by the nearly momentum-independent, non-diffusive piece of the structure factor \cite{si}.} We can quantitatively account for this behavior at $T\gtrsim J$ based on a simple (but controlled) ``local" approximation for $\chi_{\tn{spin}}''$, which is reminiscent of a dynamical mean-field theory-type approximation \cite{dmft}. Instead of ignoring the $\q-$dependence of the matrix-elements in Eq.~\ref{eqn:lehmann} altogether, we replace it by a momentum average that leads to an overall constant prefactor $\kappa$, with $\chi''_{\tn{spin}}\approx \kappa \beta\omega/J$. This is also consistent with our direct computations of $\chi''_{\tn{spin}}$ extracted from $\S(\q,\omega)$ at small $\omega$ (Fig.~\ref{fig:N2}(b) inset). We emphasize that there is a crossover out of the $\chi''_{\tn{spin}}\sim\omega$ regime and the response vanishes smoothly beyond a scale set by the spin bandwidth, which for tractability we replace with a sharp cutoff at $|\omega|=\alpha \pi J$ \cite{si}. We use the above approximate form to simplify the local electron self-energy as
\beq
 \Sigma''_{\text{loc}}(\omega,T)\approx \int d\ve~\nu(\ve)~\chi''_{\tn{spin}}(\omega-\ve)~f(\omega,\ve), 
\label{eqn:local}
\eeq
where $\nu(\ve)$ is the electronic density of states and $f(\omega,\ve)$ is as defined in Eqn.~\ref{f}. We use the simplified form of $\chi''_{\tn{spin}}$ introduced above. 

To better characterize the $\omega$ dependence of the numerically evaluated self-energy with increasing temperature, accounting for the intrinsic variations associated with $\Sigma''_{k_F}(0,T)$,  we consider the function
\beq
h(\omega,T, q_\perp,k_F) = \frac{\Sigma ''_{k_F}(q_{\perp}, \omega, T)}{\Sigma ''_{k_F}(q_{\perp}, 0, T)} - 1,
\label{eqn:freqdep}
\eeq
which trivially satisfies $h(\omega=0,T, q_\perp,k_F)=0$. We have evaluated $h(\omega,T, q_\perp,k_F)$ for a range of temperatures, $1.5J\leq T\leq 10J$, for six different $\theta\in[0,\pi/6]$ along the Fermi surface, and for the same range of $q_{\perp}$ as in Fig.~\ref{fig:N1}(c)-(d). Remarkably, we find that these curves all collapse on to a universal function, $\phi(\omega,\alpha\beta J)$ (dashed line in Fig.~\ref{fig:N2}(b)), that is computed using the local form of the spin-susceptibility in Eqn.~\ref{eqn:local}. The explicit analytical form for  $\phi(\omega,\alpha\beta J)$ appears in \cite{si}. 
Note that the only free parameter here is $\alpha$, which fixes the spin bandwidth, and can reproduce the curves for all $T\gtrsim J$ and a wide range of $\omega$; $\alpha J$ also sets the scale at which $\Sigma ''_{\text{loc}}(\omega)$ crosses over from a low frequency $\omega ^2$ behavior to the asymptotic high-frequency regime. Moreover, when $\beta \omega \rightarrow 0$ and $\beta \alpha J \ll 1$, the coefficient of this low-frequency regime scales as $\Sigma ''_{\text{loc}}(\omega)\sim \beta^4\omega^2$ (with additional dimensionful prefactors) \cite{si}. We note that the dashed line in Fig.~\ref{fig:N2}(b) captures the full $\beta-$dependence and various crossovers out of the asymptotic high-temperature regime. 

It has not escaped our attention that at large frequencies, there is a weak $\omega-$linear dependence of $\Sigma''_{k_F}(\omega)\sim\Sigma''_{\tn{loc}}(\omega)$, whose slope is independent of temperature. For the specific electronic dispersion on the triangular lattice near half-filling that is used to evaluate the self-energy, we are near a van-Hove singularity. The origin of this frequency dependence can be traced back to the electronic density of states $\nu(\omega)$, which is not independent of $\omega$ \cite{si}.

{\it Contrast with electron-phonon scattering.-}
It is useful to contrast the results obtained here for electrons scattering off a frustrated paramagnet with the more conventional example of electron-phonon scattering at $\omega_\tn{D}<T\ll \ve_{\tn{F}}$. As a function of frequency, the two problems are similar, with $\omega_\tn{D}$ playing a role analogous to the spin bandwidth. However, due to the unbounded phonon Hilbert space and an associated temperature-independent phonon spectral function, we note that, at low frequencies,  $\Sigma''_{\tn{loc}}(\omega)/\Sigma''_{\tn{el-ph}}(\omega)\sim \beta$ \cite{si}.
Similarly, as is already clear from our considerations thus far, in the high-$T$ limit and for $\omega=0$ we also find $\Sigma''_{\tn{loc}}(T)/\Sigma''_{\tn{el-ph}}(T)\sim \beta$.  
This is consistent with the classical result, whereby electrons scattering off high-temperature phonons leads to a scattering cross section that depends linearly on temperature. On the other hand, the additional suppression of the spin spectral function ($\sim\beta$) exactly cancels out this temperature dependence.

{\it Outlook.-} We have presented a quantitative theory for the electron self-energy for $J\alt T\leq\infty$ in a Fermi liquid when Kondo-coupled to a frustrated Heisenberg spin system obeying semi-classical Landau-Lifshitz dynamics. The resulting electron self-energy leads to a conundrum for the in-plane electrical transport in PdCrO$_2$, which displays a broad regime of an excess $T-$linear resistivity for $T\gtrsim J$, when compared against the iso-structural but non-magnetic compound PdCoO$_2$ \cite{Hicks15}. The distinction to PdCoO$_2$ would seem to rule out a purely electron-phonon scenario, as well as a scenario involving electrons scattering off spin-waves \cite{chernyshev}. Our present analysis disfavours
an analogous electron--local-moment scenario. Identifying the origin of this phenomenon remains a worthwhile challenge. 

{\it Acknowledgements.-} We thank E. Berg, A. Mackenzie and V. Sunko for discussions. JFMV and DC are supported by faculty startup funds at Cornell University. DC acknowledges hospitality of the Max-Planck Institute for the Physics of Complex Systems during the final stages of this work. This work was in part supported by the Deutsche Forschungsgemeinschaft under grants SFB 1143 (project-id 247310070) and the cluster of excellence ct.qmat (EXC 2147, project-id 390858490). 

\bibliographystyle{apsrev4-1_custom}
\bibliography{refs}

\clearpage
\renewcommand{\thefigure}{S\arabic{figure}}
\renewcommand{\figurename}{Supplemental Figure}
\setcounter{figure}{0}
\appendix
\pagenumbering{arabic}

\begin{widetext}

\begin{center}
  \textbf{\large SUPPLEMENTARY INFORMATION}\\[.2cm]
  \textbf{\large An intermediate-scale theory for electrons coupled to frustrated local-moments}\\[.2cm]
  Adam J. McRoberts$^{1,*}$, J.F. Mendez-Valderrama$^{2,*}$, Roderich Moessner$^{1}$, and Debanjan Chowdhury$^{2}$
  {\itshape
  	\mbox{$^{1}$Max-Planck-Institut f\"ur Physik komplexer Systeme, N\"othnitzer Stra\ss e 38, 01187 Dresden, Germany}\\
	\mbox{$^{2}$Department of Physics, Cornell University, Ithaca, New York 14853, USA.}\\
 }
\end{center}

\let\thefootnote\relax\footnote{* These authors contributed equally}

\section{Dynamical spin structure factor}

We give here a more detailed overview of the calculation and approximation of the dynamical structure factor,
\eqn{
\S(\q, \w) = \int_{-\infty}^{\infty} dt ~e^{i\w t} \avg{\bS(\q, t)\cdot \bS(-\q, 0)},
}
where the expectation value refers to the thermal average. We evaluate this correlation function at a given temperature using a combination of Monte Carlo (MC) and molecular dynamics (MD) simulations, with a linear system size $L = 120$; the number of sites is $N = L^2$. We use periodic boundary conditions in the directions specified by the chosen lattice basis, $\boldsymbol{a}_1 = (1, 0)$ and $\boldsymbol{a}_2 = (1/2, \sqrt{3}/2)$.

We construct an initial ensemble of $1000$ thermal states of the classical spin Hamiltonian $H_S$. Each state begins as a random configuration - completely independent of every other state. We then perform $N \times 10^4$ heatbath updates \cite{loison2004canonical}, where the spin on a randomly selected site is redrawn from the exact thermal distribution for a single spin in an effective magnetic field (the sum of the neighbouring spins). 

From the initial ensemble, we can calculate the \textit{static} structure factor, $\S(\q) = \avg{\bS(\q)\cdot \bS(-\q)}$, by taking the Fourier transform of each state, and calculating the ensemble average of the result.

To calculate the dynamics, we numerically integrate the classical equations of motion using the standard 4th-order Runge-Kutta method. We use a step size $\Delta t = 0.002 J^{-1}$, and evolve each state to a final time $t_f = 4096 J^{-1}$. We take $t_f$ to be sufficiently large that all correlations (in real-time) have decayed, and we are justified in approximating
\eqn{
\S(\q, \w) \approx \int_0^{t_f} dt ~e^{i\w t} \avg{\bS(\q, t)\cdot \bS(-\q, 0)} = \frac{1}{t_f} \avg{\bS(\q, \w) \cdot \bS(-\q, -\w)},
}
where the frequency Fourier transform refers to the discrete Fourier transform, over a finite time. 

The numerical simulations suffice to capture the physics of the spin Hamiltonian $H_S$, but our aim is to use the spin structure factor to compute semiclassical corrections to the electron dynamics. Here we are faced with the problem that the electron bandwidth is two orders of magnitude greater than the spin bandwidth -- and the correction to the electron self-energy (\ref{eqn:spectral_rep_sigma}) involves the convolution, $\S(\q, \w - \epsilon_{\k + \q})$, with the electron dispersion. By virtue of the discrepancy in the bandwidths, the convolved dynamical structure factor has support only over a very narrow (and $\k$-dependent) region of momentum $\q$. This precludes the numerical calculation of the electron self-energy integral, because the momentum resolution from the simulations is insufficient to facilitate its reliable evaluation.

To obtain the required resolution in momentum space, we require an analytic approximation of the dynamical structure factor.
Observing the numerical data, we find that, at temperatures $T \gtrsim J$, to a good approximation, the frequency dependence follows the phenomenological form
\eqn{
\S(\q, \w) = \frac{\S(\q)\mathcal{N}(\alpha_\q, \eta_\q)}{\sinh^2(\alpha_\q \w) + \eta_\q},
\label{ansatz}
}
where the numerator is fixed by the requirement $\int d\w \S(\q, \w) = \S(\q)$. This implies that the normalisation factor is given by
\eqn{
\mathcal{N}(\alpha, \eta) = \frac{\alpha \sqrt{\eta(\eta - 1)}}{\arcsinh(\sqrt{\eta - 1})}.
}

We now have to find analytic approximations for the three momentum-dependent functions, $\S(\q)$,  $\alpha_{\q}$, and $\eta_{\q}$. 


We begin by calculating the static structure factor. The full generating functional for the classical spins in thermal equilibrium is:
\eqn{
\Z[\J] = \int \mD \W \exp \left(-\frac{\beta J}{2} \sum_{i,\boldsymbol{e}} \bS_i \cdot \bS_{i+\boldsymbol{e}} + \sum_i \J_i^{\mu} S_i^{\mu} \right),
}
where we have expressed the classical spin Hamiltonian $H_S$ in terms of a sum over the nearest-neighbour vectors $\boldsymbol{e}$, and the measure $\mD \W$ indicates that we integrate only over configurations that satisfy the unit-length constraint. 

To make analytic progress, we use a soft-spin approximation -- relaxing the unit-length constraint, we add a Lagrange multiplier term to the effective Hamiltonian that imposes the constraint on average. We now write the measure as $\mD S$, to indicate integration over all real values of $S^x, S^y$, and $S^z$ independently. The generating functional becomes:
\eqn{
\Z[\J] = \int \mD S \exp \left(-\frac{\lambda}{2}\sum_i S_i^{\mu} S_i^{\mu} - \frac{\beta J}{2} \sum_{i,\boldsymbol{e}} \bS_i \cdot \bS_{i+\boldsymbol{e}} + \sum_i \J_i^{\mu} S_i^{\mu} \right).
}

Fourier transforming, the generating functional becomes:
\eqn{
\Z[\J] = \int \left(\prod_{\q} dS^{\mu}_{\q}\right) \exp \left(-\frac{\lambda}{2}\sum_{\q} S_{\q}^{\mu} S_{-\q}^{\mu} - \frac{\beta J}{2} \sum_{\q} \gamma(\q) S^{\mu}_{\q} S^{\mu}_{-\q} + \sum_{\q} \J_{\q}^{\mu} S_{-\q}^{\mu} \right),
}
where
\eqn{
\gamma(\q) := \sum_{\boldsymbol{e}} e^{i \bq \cdot \boldsymbol{e}} = 2\cos(q_x) + 4\cos\left(\frac{q_x}{2}\right)\cos\left(\frac{\sqrt{3}q_y}{2}\right).
}
Performing the Gaussian integrals, we obtain:
\eqn{
\log\left(\frac{\Z[\J]}{\Z[0]}\right) = \sum_{\q, \mu} \frac{\J^{\mu}_{\q} \J^{\mu}_{-\q}}{\lambda + \beta J \gamma(\q)},
}

and so the static structure factor is given by
\eqn{
\S^{\mu\nu}(\q) = \avg{S^{\mu}_{\q} S^{\nu}_{-\q}} = \frac{\delta^{\mu\nu}}{\lambda + \beta J \gamma(\q)}, \;\;\;\;\;\;\; \S(\q) = \sum_{\mu} \S^{\mu\mu}(\q) = \frac{3}{\lambda + \beta J \gamma(\q)}.
}

The Lagrange-multiplier is obtained from the self-consistency equation,
\eqn{
1 = \int_{\q} \S(\q) = \int_{\q} \frac{3}{\lambda + \beta J \gamma(\q)},
}
where $\int_{\q}$ denotes the normalised integral over the Brillouin zone (i.e., $\int_{\q} 1 = 1$). This equation is straightforward to solve numerically. The static structure factor obtained from the soft-spin approximation is in very good agreement with the results of the MC simulations at temperatures $T \gtrsim J$ (see Fig.~\ref{fig:ssf}).

\begin{figure}
    \centering
    \includegraphics{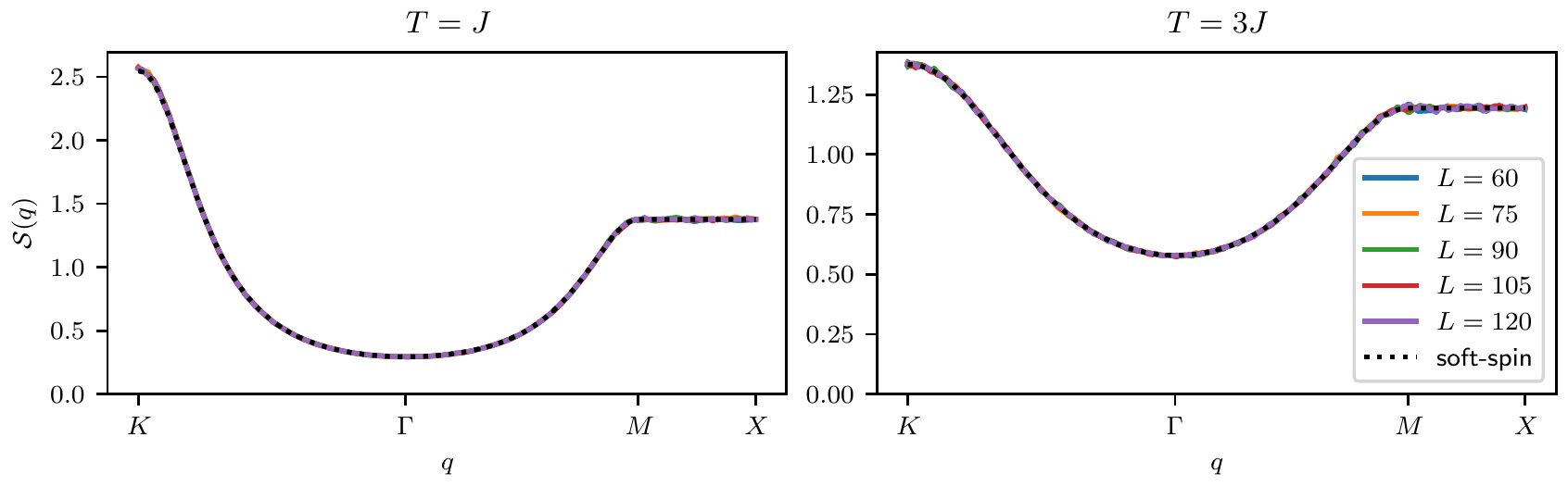}
    \caption{Comparison of the static structure factor obtained from numerical simulations at various system sizes with the soft-spin approximation, across the indicated momentum-cut between high-symmetry points in the Brillouin zone. We observe that any finite-size effects have vanished at the system sizes we consider, and that there is very good agreement between the numerical results and the analytic approximation at these temperatures.}
    \label{fig:ssf}
\end{figure}

We are left with the momentum-dependent fitting parameters $\alpha_{\q}$ and $\eta_{\q}$, which we obtain as functions of momentum by fitting the numerical data to the ansatz (\ref{ansatz}) at each discrete $\q$. To complete the approximation of $\S(\q, \w)$, we need to express $\alpha_{\q}$ and $\eta_{\q}$ as functions of momentum in terms of a non-extensive number of coefficients.

Now, $\alpha_\q,~\eta_\q$ respect the underlying space group symmetries of the triangular lattice, and can thus be expressed in terms of the following objects: 
\beq
\gamma_n(\q) = \frac{1}{|E_n|} \sum_{\vec{\delta} \in E_n} e^{i\q\cdot \vec{\delta}},
\label{eq:gn}
\eeq
where $E_n$ is the set of $n^{\tn{th}}$ nearest-neighbour vectors, and $|E_n|$ is the cardinality of that set. 

A final subtlety is that, since we have diffusion in the long-wavelength limit, we have $\eta_q \sim q^4$, $q \sim 0$. We change the leading dependence by defining $\tilde{\eta}_{\q} = \eta_{\q} / (6 - \gamma(\q))^2$. We can now express $\alpha_\q,~\eta_\q$ as
\eqa{
\alpha_{\q} &= \sum_n \alpha_n \gamma_n(\q), \nn \\
\eta_{\q} &= (6 - \gamma(\q))^2\sum_n \tilde{\eta}_n \gamma_n(\q),
\label{eq:aqhq}
}
where the coefficients are given by
\eqn{
\alpha_n = \int_{\q} \;\gamma_n(\q) \alpha_{\q}, \;\;\; \tilde{\eta}_n = \int_{\q} \;\gamma_n(\q) \tilde{\eta}_{\q},
\label{eq:anhn}
}
In the high-temperature limit with a short correlation length, the coefficients $\alpha_n$ and $\tilde{\eta}_n$ rapidly decay, and we can truncate the series (\ref{eq:aqhq}) at $n = 5$. We show the values and temperature dependence of these coefficients in Fig.~\ref{fig:alpha_eta}. Inserting these back into (\ref{ansatz}) provides an analytic approximation of the dynamical structure factor -- in very good agreement with the MD simulations for $T \gtrsim J$ -- which we may use to calculate perturbative corrections to the electron dynamics.

\begin{figure}
    \centering
    \includegraphics{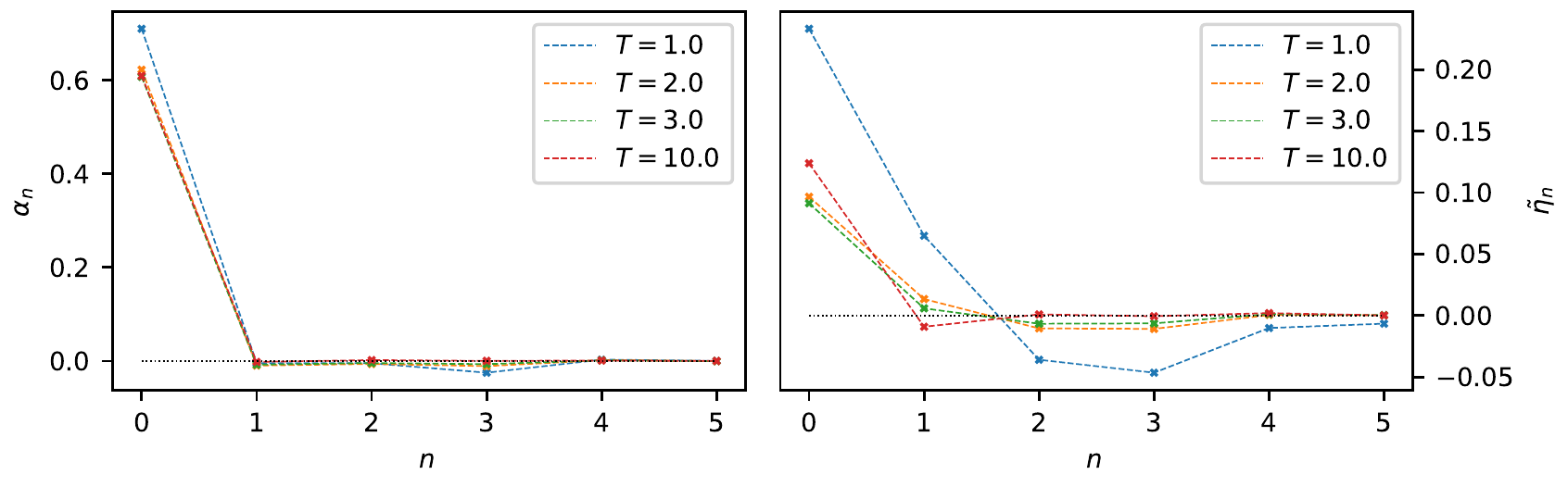}
    \caption{The values of the coefficients, $\alpha_n$ and $\tilde{\eta}_n$, up to $n = 5$, for various temperatures. We observe that the truncation at $n = 5$ is well-justified, in particular for $T \gtrsim 2J$.}
    \label{fig:alpha_eta}
\end{figure}

\section{From ``classical" structure factor to electron self-energy}

In the main text, we use the fluctuation dissipation theorem (FDT) to connect the spin susceptibility, $\chi''(\q,\omega)$, that enters Eqn.~\ref{eqn:spectral_rep_sigma} to the numerically evaluated dynamical structure factor $\mathcal{S}(\q,\omega)$ . In the particular semi-quantum regime of interest, we work with the classical limit of the FDT from the outset. This limit is taken first to preserve the analytic properties of $\chi''(\q, \omega)$. Note that since the classical $\mathcal{S}(\q,\omega)$ obtained using Eqn.~\ref{eqn:ll} does not satisfy  $\mathcal{S}(\q,-\omega)= \mathcal{S}(\q,\omega)e^{-\beta \omega}$, applying the quantum FDT instead leads to a  $\chi''(\q,\omega)$ that is not antisymmetric as a function of frequency.

\section{Details of numerics for evaluation of self-energy}

The numerical evaluation of the electron self-energy in Eq.~\ref{eqn:spectral_rep_sigma} poses technical challenges due to the smallness of the spin bandwidth (set by $J$) relative to the electronic bandwidth. For the parameters listed in the main text, the bandwidth ratio is approximately $J/W\approx 1/500$. In practice, this means that when integrating over momenta, the spin spectral function has support in a narrow range of momentum near the Fermi surface. The width of this region is determined by the condition 
\beq
|\omega-\ve_{\q+\k}|<2 \pi J .
\label{eqn:domain}
\eeq
Here $\q$ denotes the momentum variable over which the integration is performed, while $\k$ is the external momentum. Beyond this region, the contribution to the self-energy is negligible. To calculate the integral we use an adaptative domain of integration that only samples points in momentum space that fulfill the condition in Eqn.~\ref{eqn:domain}. To simplify the sampling procedure, we shift $\q\rightarrow \q-\k$ and perform the integral in polar coordinates. The integration region consists of ``rays" centered at the Fermi surface that are generated at chemical potential $\mu +\omega $. These rays extend radially to the boundaries of the region delimited by Eqn.~\ref{eqn:domain}. Importantly, the rays are generated in a way that their angular separation is kept constant. Taking into account the change in the integration measure, the radial integration is performed using Romberg's method \cite{recipes}. A condition for the implementation of this method is to sample $2^k +1$ uniformly along each axis. For the rapidly changing radial axis, we chose $k=16$ and for the smoother angular integration we chose $k=10$.

\section{Analytical evaluation of the electron self-energy in the local approximation}
In this section we provide an explicit expression for the self energy in the local approximation, which is compared against the full numerical result in the main text. The first thing to note is that, putting the explicit limits of integration that reflect the cutoff arising from the spin bandwidth, the self energy becomes:
\beq
 \Sigma''_{\text{loc}}(\omega,T)\approx  g^2\nu(\omega)\int_{\omega -\alpha  J}^{\omega+\alpha J } d\ve~\chi''_{\tn{spin}}(\omega-\ve)~f(\omega,\ve), 
\label{eqn:local2}
\eeq
where we reintroduced the coupling constant $g$ and we assumed that the density of states $\nu (\ve)$ varies slowly in the range $[\omega-\alpha J, \omega+\alpha J]$. Note that this slow variation does not necessarily imply that $\nu (\omega)\approx \nu (0)$ in the proximity of a van-Hove singularity when $\omega>\alpha J$. The integrals in Eqn.~\ref{eqn:local2} can then be performed exactly using the local form of the spin spectral function, 
\beq
 \Sigma''_{\text{loc}}(\omega,T) &=& \kappa \alpha  g^2 \nu(\omega) Y(\beta\omega, \alpha \beta J), \\
Y(x,z) &=&  -\frac{z}{2}-\log\left[\left(e^{-z-x}+1\right)\left(e^{z-x}+1\right)\right]+\frac{\text{Li}_{2}\left(-e^{-z-x}\right)-\text{Li}_{2}\left(-e^{z-x}\right)-2\text{Li}_{2}\left(1-e^{z}\right)}{z},
\label{eqn:local3}
\eeq
where $\text{Li}_2$ is the dilogarithm and $\kappa$ is defined in the main text. From this analytical result it is then inferred that the crossover scale from the low-$\omega$ to the high-$\omega$ regime is set by $\alpha J$ at all temperatures. We use this result for the local self energy to construct the function $\phi$ in the main text. In particular, applying the analogous in Eqn.~\ref{eqn:freqdep} to $\Sigma''_{loc}$ we find:
\beq
\phi( \omega,\alpha \beta J)=\frac {\nu(\omega)}{\nu(0)}\frac{Y(\beta \omega,\alpha \beta J)}{Y(0,\alpha \beta J)}-1,
\label{eqn:scalingF}
\eeq
that we then use to generate the collapse in Fig.~\ref{fig:N2}.

Additionally, in the high$-T$ limit $\beta J\rightarrow 0$, $Y(\beta\omega, \alpha\beta J)\rightarrow 2$ regardless of the frequency. This behavior matches the numerical result that the self-energy becomes frequency independent at high temperatures. In this regime, the only residual frequency dependence originates from $\nu(\omega)$ which has a slow but non-negligible effect as detailed in the main text. Additionally, we can determine the low frequency behavior by expanding near $\omega=0$, which yields
\beq
 \Sigma''_{\text{loc}}(\omega,T) -  \Sigma''_{\text{loc}}(0,T) \approx \kappa \alpha  g^2 \nu(0) \frac{\left(\sinh\left(\beta\alpha J\right)-\beta\alpha J\right)\text{sech}^{2}\left(\frac{\beta\alpha J}{2}\right)}{4\alpha\beta J}\left(\beta\omega\right)^{2}. 
\eeq
At high temperatures, we can expand the hyperbolic functions, and we find that $\Sigma''_{\text{loc}}(\omega,T) -  \Sigma''_{\text{loc}}(0,T) \approx \kappa \alpha^3  g^2 \nu(0) (\beta J)^2 (\beta\omega)^2$ as detailed in the main text.

\section{Filling dependence of electron self-energy}

\begin{figure*}[htbp!]
\centering
\hspace*{-0.9cm} 
\includegraphics[width=0.85\textwidth]{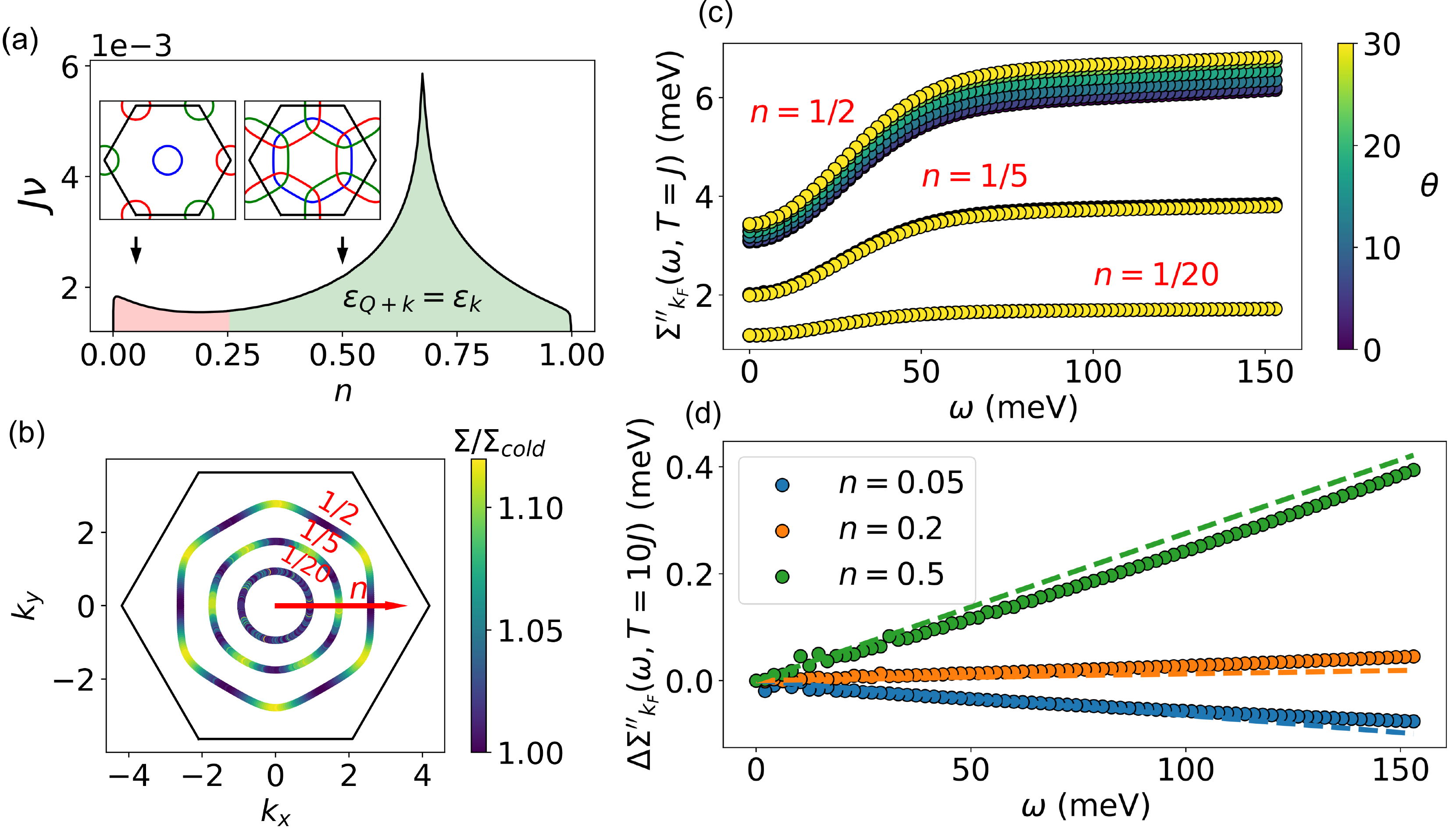}
\caption{(a) . Density of states as a function of filling. The green region denotes the fillings for which hot spots can emerge for our particular fermiology. b) Ratio of the self energy to the self energy at the 'coldest' points in the Fermi surface calculated at different fillings. This shows a remanant signal at low fillings below the threshold in (a) due to thermal broadening. c-d) Frequency dependence of the self energy at different fillings. c) $T=J$ for different $\theta$ along the Fermi surface. d) subtracting the $\omega=0$ contribution at $T=10J$. } 
\label{fig:Napp1}
\end{figure*}

In the main text, we noted that short-ranged spin fluctuations remain strong at $T\sim J$ near $\vec{Q} \in \{ \vec{K}, \vec{K}' \}$.  As a consequence,  $\Sigma''(k_F,0, T=J)$ displayed a series of hot-spots that signal enhanced scattering with momentum transfer  $\vec{Q}$ for which the condition 
\beq
\ve_{\vec{Q}+\k}=\ve_\k
\label{eqn:hot}
\eeq
is attained. Here, we study the relation between the location of these hot-spots and the filling, $n$, of the electron band. We take the convention that a fully filled band (including spin) corresponds to $n=1$. The non-interacting density of states for the specific $t,t'-$dispersion is shown in Fig~\ref{fig:Napp1}a). It is natural to expect that the size and shape of the Fermi surface determines the set of points for which scattering by $\vec{Q}$ produces an enhancement in $\Sigma''(k_F,0, T=J)$. 
With increasing $n$, the Fermi surface first fulfills the condition Eqn.~\ref{eqn:hot} when $n=0.26$ at six points along the $\Gamma -K$ and $\Gamma -K'$ lines. As the Fermi surface grows with increasing $n$, each hot-spot splits into two, generating twelve hot-spots in total. These merge again at $n=0.34$ along the $\Gamma -M$ lines before splitting once more for $n>0.34$. In the main text, we fixed $n=0.5$ inspired by the magnetic delafossite PdCrO$_2$. 

We note that substantial thermal broadening occurs at $T\sim J$ in the structure factor. This broadening is sufficient to prevent the identification of each individual hot-spot at filling $n=0.5$. Even upon going below the regime where Eqn~\ref{eqn:hot} holds, at filling $n=0.2$, the thermal broadening leads to appreciable enhancement of the self-energy along the $\Gamma-K$ and $\Gamma-K'$ lines, see Fig.~\ref{fig:Napp1}b). As the filling is further reduced, the relative enhancement of the self-energy decreases smoothly and at $n=0.1$, it is not possible to distinguish any feature from the uniform background.

The frequency dependence of $\Sigma''$ is qualitatively similar for different fillings, see Fig.~\ref{fig:Napp1}c. This is expected when  $\chi''(\q,\omega)$ does not have a strong momentum dependence. The location of the crossover from a quadratic to a slowly varying regime is the same for all fillings. This is consistent with the fact that the crossover is set by the bandwidth of the spins. In the main text, we showed that the scale of $\Sigma''$ is modulated by $\Sigma''_{k_F}(0,T)$ as the angle along the Fermi surface, $\theta$, varies. Since the hot-spots at $\omega=0$ disappear gradually with decreasing filling,  $\Sigma''_{k_F}(\omega,T)$ becomes independent of $\theta$ at $T\sim J$ and $n\ll 0.26$.

In the main text, we argued that the variation in the density of states is the main source of frequency dependence above the crossover set by $\alpha \pi J$. We can verify this for different fillings at high temperature where the local approximation is justified. In this regime, we calculate $\Delta \Sigma ''=\Sigma ''_{k_F}(\omega, T)-\Sigma ''_{k_F}(0, T)$; the results are shown in Fig.~\ref{fig:Napp1} d). Note that the variation quantified by $\Delta \Sigma''$ is an order of magnitude smaller than the bare value which is of order 4-6 meV. Then, if the only source of frequency dependence is due to the density of states, we expect at low frequencies that \beq \Delta \Sigma ''_{k_F} (\omega)=\omega  \left.\frac{d\nu}{d\omega}\right\vert_{\omega=0} \Sigma ''_{k_F}(0)/ \nu(0). \label{eqn:linear_apfil}\eeq We compare the full calculation of $\Delta \Sigma ''$ with the r.h.s of Eqn.~\ref{eqn:linear_apfil} represented by the dashed lines in Fig.~\ref{fig:Napp1} d). The agreement between these two calculations, together with the fact that the slope is temperature independent, supports the claim that the slow variation in $\Sigma ''$ is generated by changes in the density of states at high frequencies. Note that the density of states decreases rapidly at low $n$, see Fig.~\ref{eqn:hot}a). Even though these fillings are far from the Van-hove singularity, the change in $\nu$ is sufficient to leave an imprint in $\Delta \Sigma ''$, which becomes negative. 

\section{Local approximation to the structure factor}

\begin{figure*}[htbp!]
\centering
\hspace*{-0.9cm} 
\includegraphics[width=0.85\textwidth]{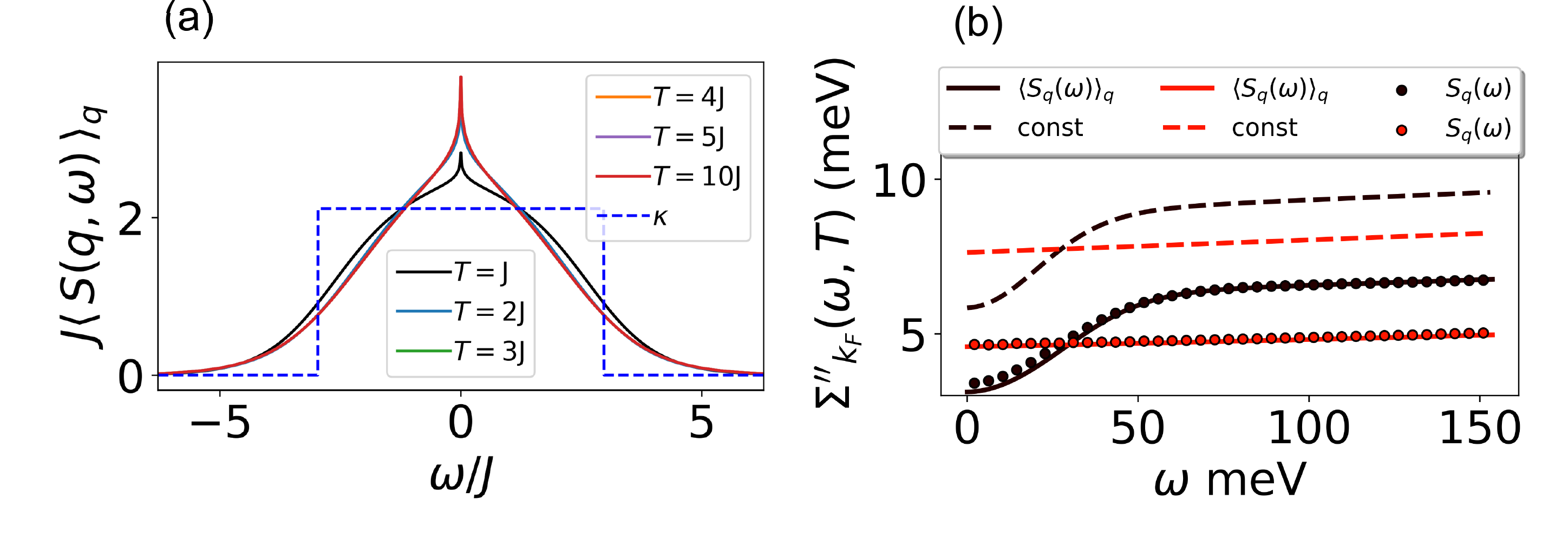}
\caption{(a) . Spin susceptibility and structure factor (inset) averaged over momenta. (b) Comparison between different levels of approximation to the self energy at $T=J$ and $T=10J$. } 
\label{fig:Napp2}
\end{figure*}

In this section, we elaborate on details of the parameters of the simplified model for $\chi ''$. As described in the main text, we have two free parameters, $\alpha$ which is related to the spin bandwidth, and $\kappa$ which is extracted from the momentum averaged dynamical structure factor.  At first glance, the value of  $\kappa$ is inconsequential for the scaling shown in  Eqn.~\ref{fig:N2} as it drops out upon calculating Eqn.~\ref{eqn:freqdep}. Using the least squares procedure, we fit $h(\omega ,T, q_{\perp}, k_F)$ in Eqn.~\ref{eqn:freqdep} to Eqn.~\ref{eqn:scalingF}, and we find $\alpha=0.94 \pm 0.05$ without ever determining the value of $\kappa$.  With this fit we are able to fix the behavior of the scaling function in Eqn.~\ref{eqn:scalingF} at all temperatures $T\geq 2J$ with good accuracy as shown in Fig.~\ref{fig:N2}. However, there is a caveat in the above procedure since both $\alpha$ and $\kappa$ are related by a sum rule:
\beq
\lim _{T\rightarrow\infty} \int d\omega \langle\mathcal{S}(\q,\omega)\rangle_\q = 4\pi
\label{eqn:sumrule}
\eeq
where $\langle \cdot \rangle_\q$ denotes an average over the first Brillouin zone. This sum rule imposes the constraint $\alpha \kappa =2$. Using this, we infer a value of $\kappa=2.1\pm 0.1$. We use these values of $\alpha$ and $\kappa$ for the inset in Fig.~\ref{fig:N2}. 

Now we can also obtain $\alpha$ and $\kappa$ independently by first fitting the momentum averaged susceptibility obtained by the full spin dynamics, $T \langle \chi ''\rangle_\q$ near $\omega =0$ at $T=100J$. With this procedure, we are able to extract a value of $\kappa =2.120\pm 0.008$ (fitting in the range ($[-J\pi/2,J\pi/2]$), where the uncertainty given by the standard error of the linear regression). Using the sum rule,  we then infer a value of $\alpha =0.943\pm 0.008 $. Therefore, the parameters extracted by fitting the Eqn.~\ref{eqn:scalingF} and the ones extracted from $T \langle \chi ''\rangle_\q$ are broadly consistent with each other.

Some of the details that are overlooked by the linear approximation to $\chi''$ are relevant for a more precise comparison with the numerical data.  Importantly, the linear approximation overestimates the maximum of $\chi''$ and the frequency at which the maximum occurs.  As a result, the self energy is consistently overestimated by the linear approximation. This overestimation is not reflected in the collapse shown in Fig.~\ref{fig:N2} by the definition of $h(\omega, T)$. However, we can improve upon our simple model by using directly $\langle \chi ''\rangle_\q$ in Eqn.~\ref{eqn:local}. A comparison between the two local approximations and the full numerical calculation is shown in Fig.~\ref{fig:Napp2}. The improved local approximation almost perfectly describes the full numerical calculation (Although it provides less analytical insight). We then conclude that the momentum dependence of $S(\q,\omega)$ only contributes to the hot-spot physics at low $T$ but otherwise has no imprint in the self energy.

\end{widetext}

\end{document}